**Title:** Investigating heterogeneous PSMA ligand uptake inside parotid glands

Caleb M Sample[1,2]

Carlos Uribe[3,4, 5]

Arman Rahmim[1,3,5]

François Benard[3,4, 6]

Jonn Wu[7,8]

Haley Clark[1,2,8]

1 Department of Physics and Astronomy, Faculty of Science, University of British Columbia, Vancouver, BC, CA

[2] Department of Medical Physics, BC Cancer, Surrey, BC, CA

[3] Department of Radiology, Faculty of Medicine, University of British Columbia, Vancouver, BC, CA

[4] Department of Functional Imaging, BC Cancer, Vancouver, BC, CA

[5] Department of Integrative Oncology, BC Cancer Research Institute, Vancouver, CA

[6] Department of Molecular Oncology, BC Cancer, Vancouver, BC, CA

[7] Department of Radiation Oncology, BC Cancer, Vancouver, BC, CA

[8] Department of Surgery, Faculty of Medicine, University of British Columbia, Vancouver, BC, CA

**Corresponding Author:**

Caleb Sample (ORCID 0000-0001-6166-4917)

csample@phas.ubc.ca

(1)-250-640-7948

411 8th St, #205

New Westminster BC, V3M3R5

**Declaration of Interest:** None

**Abstract:**

The purpose was to investigate the spatial heterogeneity of prostate-specific membrane antigen (PSMA) positron emission tomography (PET) uptake within parotid glands. We aim to quantify patterns in well-defined regions to facilitate further investigations. Furthermore, we investigate whether uptake is correlated with computed tomography (CT) texture features. Methods: Parotid glands from [18F]DCFPyL PSMA PET/CT images of 30 prostate cancer patients were analyzed. Uptake patterns were assessed using thresholding to define "high-uptake" regions, and by computing statistics within various divisions. Spearman's rank correlation coefficient was calculated between PSMA PET uptake and feature values of a Grey Level Run Length Matrix using a long and short run length emphasis (GLRLML and GLRLMS) in sub-regions of the parotid gland. Results: PSMA PET uptake was significantly higher ($p < 0.001$) in lateral/posterior regions of the glands than anterior/medial regions. Maximum uptake was found in the lateral half of parotid glands in 50 out of 60 glands. The difference in SUVmean between parotid halves is greatest when parotids are divided by a plane separating the anterior/medial and posterior/lateral halves symmetrically (out of 120 bisections tested). PSMA PET uptake was significantly correlated with CT GLRLML ($p<0.001$), and anti-correlated with CT GLRLMS ($p<0.001$). Conclusion: Uptake of PSMA PET is heterogeneous within parotid glands, with uptake biased towards lateral/posterior regions. Uptake patterns within parotid glands were found to be strongly correlated with CT texture features, suggesting the possible future use of CT texture features as a proxy for inferring PSMA PET uptake in salivary glands.

## 1. Introduction

Prostate-specific membrane antigen (PSMA) positron emission tomography (PET) is an imaging procedure primarily used for detecting prostate cancer [1], which quantifies the expression of the PSMA, which is found on prostate cancer cells, using radiolabeled PSMA ligands. These PSMA ligands also accumulate in the major salivary and lacrimal glands [2, 3, 4], by a process postulated to be at-least partially unrelated to PSMA-mediated uptake [5, 6]. While salivary gland uptake is undesirable for the purposes of radioligand therapy for prostate cancer patients, it renders PSMA PET a potentially useful quantitative imaging modality for salivary glands.

Klein-Nulent et al. [7] hypothesized that uptake of the PSMA ligand in salivary glands is associated with the functional capacity of the glands, based on two findings. First, they noted that xerostomia (subjective dry mouth) is a common side effect of $^{177}$Lu-PSMA treatment, which may be due to the uptake and cell loss in functional regions of the gland. Secondly, $^{68}$Ga-PSMA uptake is significantly lower in irradiated submandibular glands than normal glands [8, 9]. In another study, supporting PSMA PET's utility for assessing salivary gland functionality, Zhao et al. [10] showed that $^{68}$Ga-PSMA-11 PET/CT is an effective supplement to salivary gland scintigraphy (SGS), which is a method of clinically evaluating salivary gland function. Furthermore, prior $^{131}$I-radionuclide therapy has been found to significantly decrease PSMA PET uptake in salivary glands, with a high degree of inter- and intra-patient variability [11].

Assuming salivary gland functionality and PSMA PET uptake in the glands are correlated, then in principle, both inter-and intra-patient variability in uptake could provide valuable insight into the functional anatomy of the glands. Several independent analyses have shown head-and neck cancer patients to have variable intra-parotid therapeutic-dose responses during radiotherapy [12, 13, 14, 15, 16, 17, 18]. It remains unclear whether PSMA PET / computed tomography (CT) has utility for assessing intra-parotid gland anatomical variability for the purposes of radiotherapy planning.

Intra-patient PSMA ligand uptake in salivary glands has been reported as homogeneous [7]. This is a reasonable qualitative description of uptake in the glands, as they tend to appear visually homogeneous when viewed with a default window/level for a full-body PET/CT image. However, upon viewing several PSMA PET images of prostate cancer patients with varying windows and levels, we were able to detect an asymmetric trend for uptake in the parotid glands, where regions of high uptake were biased towards the lateral and posterior portions of the gland. Uptake in the submandibular glands did appear to be homogeneous.

In this retrospective study of [18F]DCFPyL PSMA PET images of 30 prostate cancer patients, we first seek to assess the spatial heterogeneity of uptake within the parotid glands, with the motive of quantifying novel trends with future applications in improving parotid gland radiotherapy dose constraints. To make our findings practically useful, we also test for a correlation between intra-parotid PSMA PET uptake and CT radiomic texture features to assess whether CT could potentially be used as a low-cost proxy when PSMA PET imaging is not available.

## 2. Materials and Methods

### 2.1 Cohort and Image Acquisition

This retrospective study was approved by an institutional review board. Full-body [18F]DCFPyL PSMA PET/CT images were de-identified for 30 previous prostate cancer patients who had previously consented to the use of their data for research studies (Mean Age 68, Age Range 45-81; mean weight: 90 kg, weight range 52 kg-128 kg). Patients were scanned, two hours following intravenous injection, from the thighs to the top of the skull on a GE Discovery MI (DMI) scanner. PET images were reconstructed using VPFXS (OSEM with TOF and PSF corrections) (pixel spacing: 2.73-3.16 mm, slice thickness: 2.8-3.02 mm). Helical CT scans were acquired on the same scanner (kVP: 120, pixel spacing: 0.98 mm, slice thickness: 3.75 mm). To avoid imposing interpolation methods, voxel sizes were not resampled.

### 2.2 Parotid Gland Delineation:

CT images were used for delineating parotid and submandibular glands. This consisted of two steps, where Limbus AI [19] was first used for auto-segmentation of the glands, which were then manually refined by a single senior radiation oncologist, XX. Unless stated otherwise, all reported results are defined within these CT-defined regions.

## 2.3 Analysis

### 2.3.1 Parotid Gland PSMA PET Uptake

The analysis was performed using custom software written in the Python programming language.

Standard Uptake Values (SUVs) in PET Images were normalized by lean body mass ($SUV_{lbm}$) where lean body mass was estimated for each patient using the Hume formula [20]. For the remainder of this article, for brevity, SUV should be interpreted as $SUV_{lbm}$. Both $SUV_{mean}$ and $SUV_{max}$ were included in the results as relevant metrics inside various regions of interest (ROIs). For the investigation of intra-gland uptake, right and left gland statistics were pooled whilst respecting the central axis of symmetry, meaning that a "medial-lateral" direction was used rather than simply "left-right".

In the first part of the analysis, we determine typical SUV cohort statistics within the parotid and submandibular glands, as well as within the superficial and deep lobes of the parotid gland, with the latter not previously reported in the literature. Superficial and deep lobes were contoured by dividing the whole-parotid gland contours according to the standard procedure outlined by Zhang et al. [21].

To visualize the extent of population-level spatial uptake asymmetry along the medial-lateral and anterior-posterior directions in the parotid glands and symmetry in the submandibular glands, as seen in our initial patient-specific inspection, PSMA PET SUVs for each gland type were plotted as a function of lateral and posterior displacement from the gland's center of mass. This displacement was recorded as a fraction of the gland's maximum half width in each direction (medial-lateral or anterior-posterior). Sturges' rule using the average number of voxels within glands was used to determine the optimal number of displacement bins for creating a histogram of uptake vs displacement ($number\ of\ bins = \log_2(number\ of\ voxels) + 1$).

The $SUV_{mean}$ for each patient's left and right parotid was found in each displacement bin, and finally, $SUV_{mean}$ was averaged over all glands and patients, $\overline{SUV_{mean}}$ , in each bin.

We then analyzed how regions of high PSMA PET uptake in the parotid gland are distributed, using various lower thresholds for defining said regions. Thresholds were chosen as combinations of $SUV_{mean}$ ($\mu$) and the standard deviation, $\sigma$, within the whole gland, The average displacement of various high-uptake regions from the whole parotid gland center of mass is reported.

Next, we perform a sub-region analysis where the parotid gland is divided into four sub-segments of equal volume, separately, along each Cartesian direction. Sub-segmentation is performed with parallel planar cuts, such that the 'slab volume' (total planar area multiplied by the image slice thickness) [18] is equal within an error tolerance of 0.01% in each region. $\overline{SUV_{mean}}$ and $\overline{SUV_{max}}$ (averaged over all patients) were computed in each sub-region.

Optimal planes for dividing the parotid gland in half for both maximizing and minimizing the difference between $\overline{SUV_{mean}}$ and $\overline{SUV_{max}}$ between halves was determined. Planes were defined by their normal vectors in spherical coordinates, with the zenith direction corresponding to the patient's superior direction, and the azimuthal angle of 0$^{o}$ being defined in the lateral direction and increasing towards the posterior direction. Azimuthal and polar angle combinations were used to define the planes, using azimuthal angles between 0 and 165 degrees (increments of 15 degrees) and polar angles between 0 and 90 degrees (increments of 10 degrees). This resulted in a total of 120 cutting planes being tested. The dividing plane always passed through the parotid gland's center of mass, and uptake statistics were calculated separately for voxels below and above the plane.

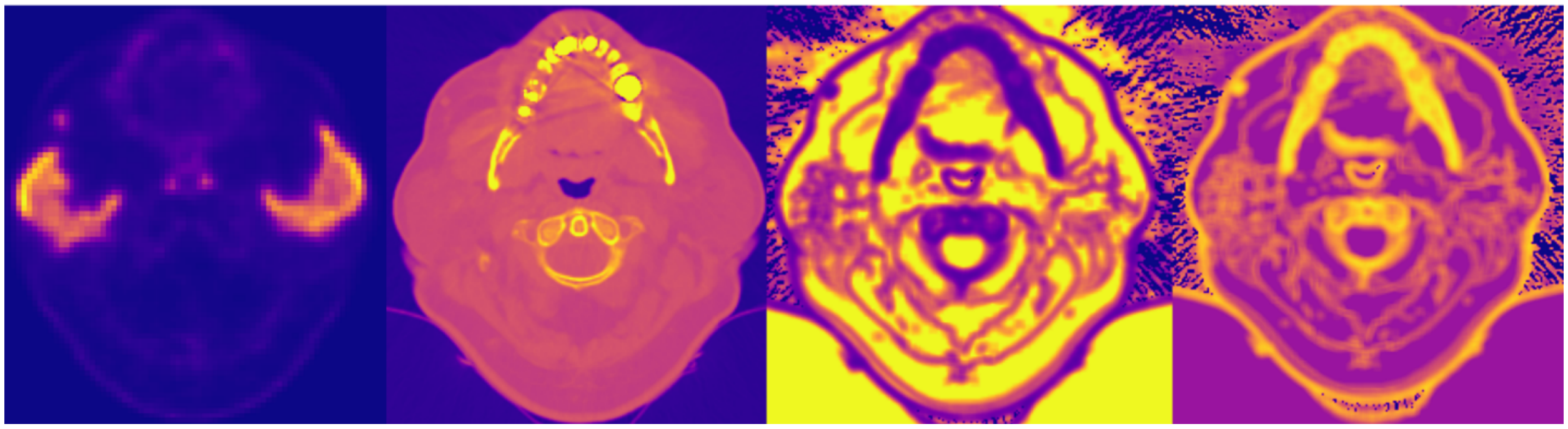

**Fig. 1** Axial slices of PSMA PET, CT, and two CT texture features, the gray level run length matrix with a long- and short-run-emphasis (GLRLML and GLRLMS) are shown through the parotid glands.

### 2.3.2 Comparison of Intra-Parotid Uptake Trends with CT Texture Features

We evaluated the correlation between PSMA PET uptake in various sub-regions of the parotid glands with regular CT Hounsfield units (HUs) as well as two CT texture feature map images. Texture features computed included the Grey-Level-Run-Length-Matrix (GLRLM) with a short (GLRLMS) and long (GLRLML) run length emphasis. Only two texture features were evaluated to reduce the likelihood of false discoveries from multiple feature testing. An axial slice containing the parotid glands is shown for PSMA PET, GLRLML, GLRLMS, and regular CT in Fig. 1.

The GLRLM was chosen for this study as it is a standard texture metric, but there are many other features which could be tested. The GLRLM is a method of extracting higher order statistical texture features from images [22], which quantifies the distribution of consecutive voxels having the same discretized grey level. The GLRLMS weights runs by the inverse square of the run length, while the GLRLML weights runs by the square of the run length. We chose to calculate both long-run and short-run versions of the GLRLM because we were unsure which quantity would be more suitable for comparing with PSMA PET.

Voxel-based features maps were calculated using the pyradiomics library [23] with a 5x5 masked kernel (voxels outside parotids not included in run lengths). Grey levels were discretized according to recommended best practices [24] into a number of bins determined using Sturges' rule with the average number of voxels in each gland. Texture features were compressed to a single scalar value for each sub-region by averaging over all directions and voxels. Feature maps were calculated on regular CTs after constraining their maximum values to 1000 HU, to lower the effect of artifacts on grey-level discretization.

The correlation of PSMA PET uptake with CT images and texture features was tested between $\overline{SUV_{mean}}$, CT HUs, GLRLML and GLRLMS corresponding to values calculated in the 12 sub-regions considered in the sub-region analysis as well as the superficial and deep lobes, making a total of 14 sub-regions. Values in sub-regions were normalized as fractional deviations from the whole-gland mean. The Spearman's rank correlation coefficient, $r_s$ and its corresponding p-value was calculated between PSMA PET-uptake and CT/CT texture features. Lastly, a Benjamini–Hockberg (BH) false discovery correction was performed, as outlined by Rahmim et al. [25].

Correlations were then re-tested using PSMA PET uptake and CT texture features within 18 non-overlapping, equal-volume sub-regions of parotid glands, as defined by Clark et al. [18]. Correlations between both normalized and absolute sub-regional values were computed. Lastly, we tested the correlation between whole-gland $SUV_{mean}$, GLRLML and GLRLMS.

**3 Results:**

PSMA PET uptake statistics are reported in Table 1 for parotid and submandibular glands, as well as the superficial and deep lobes of the parotid gland. Whole gland statistics agreed with those reported in the literature [7, 26]. $\overline{SUV_{mean}}$ and $\overline{SUV_{max}}$ in the superficial lobe were significantly higher than in the deep lobe ($p < 0.001$). $\overline{SUV_{mean}}$ in submandibular glands appeared to be mostly homogeneous, while parotid glands had increased $\overline{SUV_{mean}}$ and $\overline{SUV_{max}}$ in posterior and lateral regions ($p < 0.001$) as shown in Fig. 2.

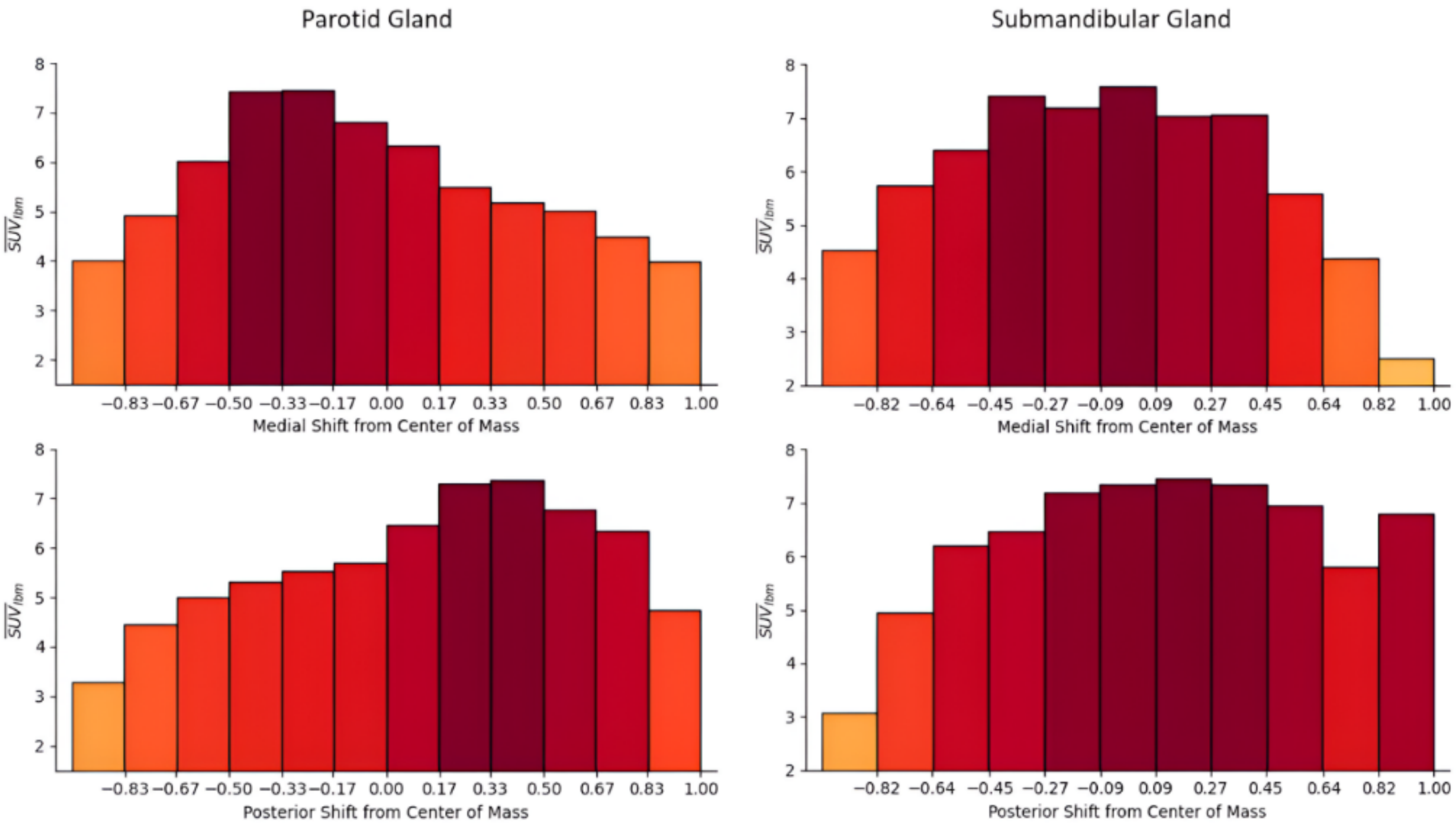

**Fig. 2** PSMA PET $\overline{SUV_{mean}}$ in regions shifted along the medial-lateral (top) and anterior-posterior (bottom) directions in the parotid (left) and submandibular (right) glands. Displacements were recorded as fractions of the maximum half width of the gland in the appropriate direction (medial-lateral or anterior-posterior).

**TABLE 1.** Population-level PSMA PET uptake statistics (n=60) are shown for the parotid and submandibular glands, as well as the superficial and deep lobes of the parotid gland.

| Region of Interest | | $\overline{SUV_{mean}}$ | $\overline{SUV_{max}}$ |
|---|---|---|---|
| Parotid Gland | Whole | $6.2 \pm 1.7$ | $15.5 \pm 4.5$ |
| | Left | $6.2 \pm 1.7$ | $15.7 \pm 4.5$ |
| | Right | $6.1 \pm 1.7$ | $15.3 \pm 4.5$ |
| | Deep Lobe | $5.0 \pm 1.7$ | $12.7 \pm 3.3$ |
| | Superficial Lobe | $6.5 \pm 1.9$ | $15.4 \pm 4.4$ |
| Submandibular Gland | Whole | $6.5 \pm 1.9$ | $15.3 \pm 3.3$ |
| | Left | $6.5 \pm 1.9$ | $15.1 \pm 3.7$ |
| | Right | $6.6 \pm 1.8$ | $15.4 \pm 3.8$ |

**TABLE 2.** Shift from the whole gland's center of mass, for regions of high PSMA PET uptake. Shifts are listed as fractions of the half width along the respective patient axis. For example. A shift of 0 means the high uptake region's center of mass is in perfect alignment with that of the whole gland, while a value of 1 means the center of mass is shifted to the furthest edge of the gland. High-uptake regions are defined with a lower threshold in terms of $\mu$ and $\sigma$, the mean and standard deviation within the whole parotid.

| High-Uptake Region Threshold | Center of Mass Shift (fraction of half width) | | |
|---|---|---|---|
| | Lateral | Posterior | Superior |
| $\mu$ | $0.06 \pm 0.09$ | $0.08 \pm 0.09$ | $0.03 \pm 0.07$ |
| $\mu + \sigma$ | $0.22 \pm 0.13$ | $0.11 \pm 0.11$ | $0.03 \pm 0.10$ |
| $\mu + 3\sigma/2$ | $0.33 \pm 0.15$ | $0.10 \pm 0.16$ | $0.04 \pm 0.18$ |
| $\mu + 2\sigma$ | $0.41 \pm 0.24$ | $0.03 \pm 0.30$ | $0.08 \pm 0.30$ |

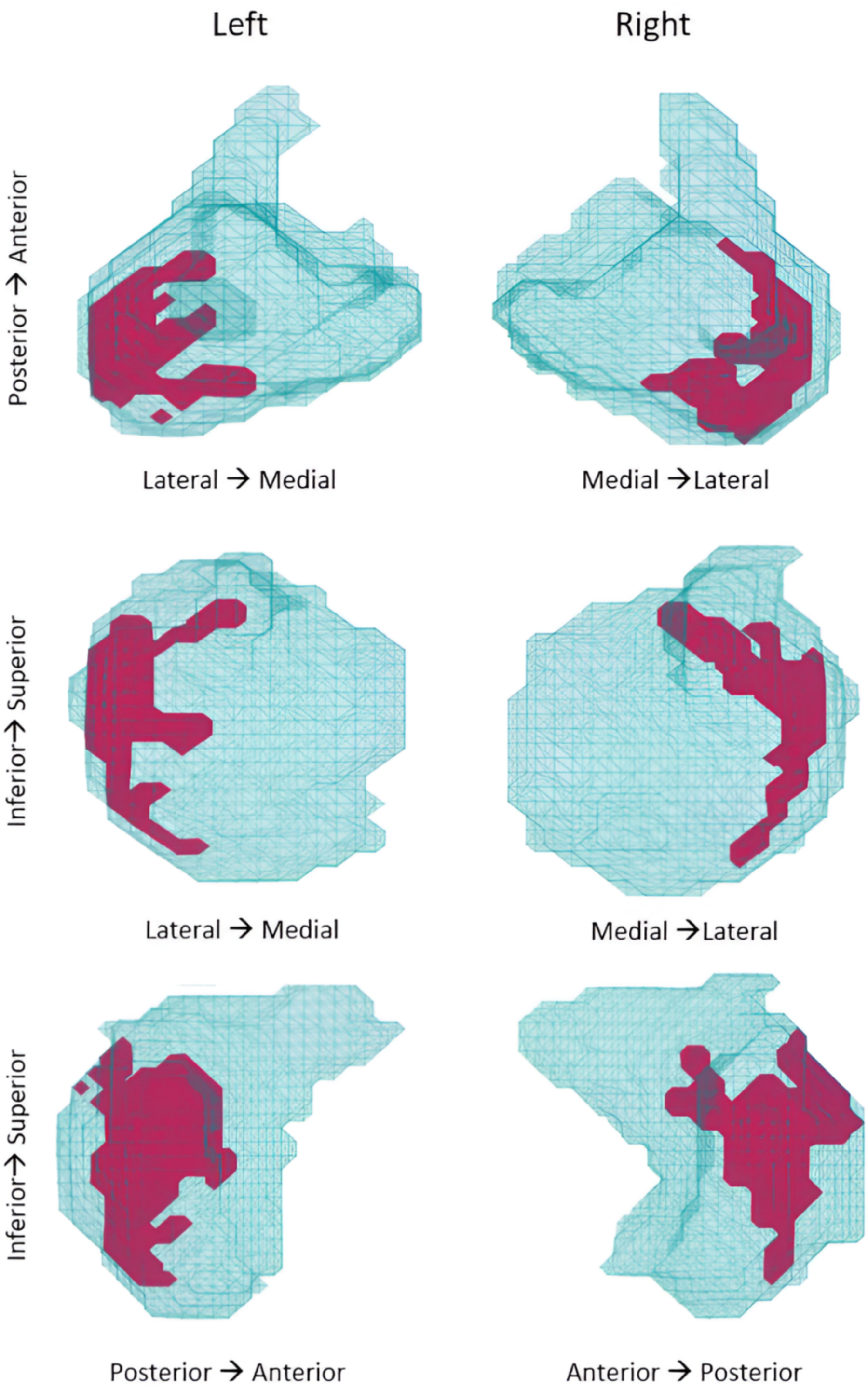

**Fig. 3** A sub-region of high PSMA PET uptake (pink), delineated using a lower threshold of $\mu + \frac{3}{2}\sigma$ inside the left and right parotid (cyan) for a representative patient. $\mu$ and $\sigma$ are the mean and standard deviation of uptake in the whole parotid.

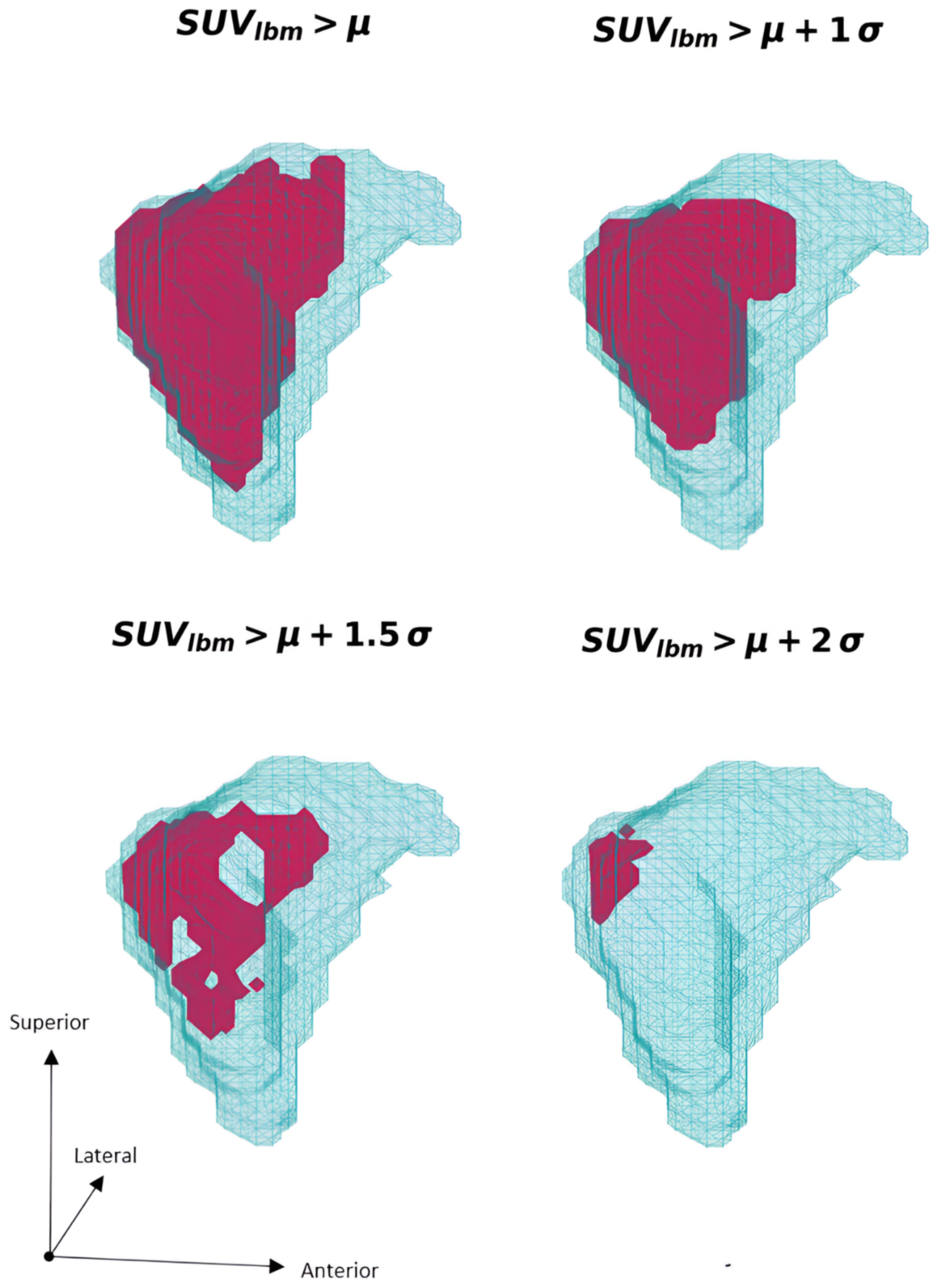

**Fig. 4** Variability of high-uptake sub-regions vs threshold for a representative patient. Thresholds are defined in terms of $\mu$ and $\sigma$, the mean and standard deviation of uptake in the whole parotid gland. The region of highest uptake tends towards the lateral and posterior (and somewhat superior) portions of the gland.

Uptake statistics for sub-regions defined by dividing the parotid into four equal-volume regions along each Cartesian direction are shown in Table 3 and visualized in Fig. 5. $\overline{SUV_{mean}}$ varied greatly between sub-regions, with regions of highest uptake for the three cutting planes being the two lateral-most sub-regions for the sagittal plane, the posterior sub-region for the coronal plane, and the middle-superior region for the transverse plane. $\overline{SUV_{mean}}$ in the two lateral-most sub-regions of the gland differed insignificantly from each other (paired $t(59) = 0.14$, $p > 0.88$) but were 45 % higher than the medial-most sub-region (paired $t(59) = 8.0$, $p < 0.001$). Out of all 60 parotids examined, 50 $SUV_{max}$ were found in the lateral half of the gland, with 30 found in the lateral-most sub-region. $\overline{SUV_{mean}}$ in the posterior-most sub-region of the gland was 42% higher than the anterior-most sub-region (paired $t(59) = 7.5$, $p < 0.001$). $\overline{SUV_{mean}}$ was highest in the two interior sub-regions when cut transversely, with the middle-superior sub-region being 6% larger than the middle-inferior sub-region ($t(59) = 5.3$, $p < 0.001$). The $\overline{SUV_{max}}$ varied with the same trend as the $\overline{SUV_{mean}}$ for the sagittal and transverse cutting planes but was found to be much more homogeneous than $\overline{SUV_{mean}}$ for the coronal cutting plane.

The optimal dividing planes for the parotid gland to alternatively maximize and minimize the difference between $\overline{SUV_{mean}}$ and $\overline{SUV_{max}}$ in each half are summarized in Table 4. The results are visualized for the maximum and minimum separation of $\overline{SUV_{mean}}$ in Fig. 6. The maximum separation of $\overline{SUV_{mean}}$ occurred with a plane vector defined by $(\phi, \theta) = (135^{\circ}, 90^{0})$, which is a diagonal split separating the posterior-lateral and anterior-medial regions of the parotid glands. In this division, the posterior-lateral portion's $\overline{SUV_{mean}}$ was 41% higher than the anterior-medial region (paired $t(59) = 9.5$, $p < 0.001$. The maximum separation of $\overline{SUV_{max}}$ occurred with a similar plane vector, $(\phi, \theta) = (165^{\circ}, 80^{0})$, with the lateral-most half's $\overline{SUV_{max}}$ being 12% higher (paired $t(59) = 7.7$, $p < 0.001$).

**TABLE 3.** PSMA PET uptake statistics. Statistics are shown in parotid gland sub-regions created by dividing glands with parallel planes into four equal-volume regions. This is done separately with sagittal, coronal, and transverse cutting planes. The final column includes the number of times the whole gland's maximum uptake voxel was found in the respective region.

| Cutting Plane | Sub-region | $\overline{SUV_{mean}}$ | $\overline{SUV_{max}}$ | Location of $SUV_{max}^{whole}$ ( n=60) |
|---|---|---|---|---|
| Sagittal | Lateral | $7.1 \pm 2.3$ | $15.2 \pm 4.5$ | 30 |
| | Middle-Lateral | $7.1 \pm 2.1$ | $14.7 \pm 4.2$ | 20 |
| | Middle-Medial | $6.1 \pm 1.8$ | $13.9 \pm 3.8$ | 5 |
| | Medial | $4.9 \pm 1.5$ | $12.9 \pm 3.3$ | 5 |
| Coronal | Posterior | $7.2 \pm 2.3$ | $14.7 \pm 4.2$ | 13 |
| | Middle-Posterior | $6.8 \pm 1.9$ | $14.9 \pm 4.2$ | 18 |
| | Middle-Anterior | $6.0 \pm 1.6$ | $14.6 \pm 4.2$ | 12 |
| | Anterior | $5.1 \pm 1.8$ | $14.4 \pm 4.5$ | 17 |
| Transverse | Inferior | $5.4 \pm 1.6$ | $14.0 \pm 4.2$ | 9 |
| | Middle-Inferior | $6.7 \pm 1.9$ | $14.7 \pm 4.3$ | 13 |
| | Middle-Superior | $7.1 \pm 2.1$ | $14.9 \pm 4.5$ | 20 |
| | Superior | $5.9 \pm 1.8$ | $14.8 \pm 4.4$ | 18 |

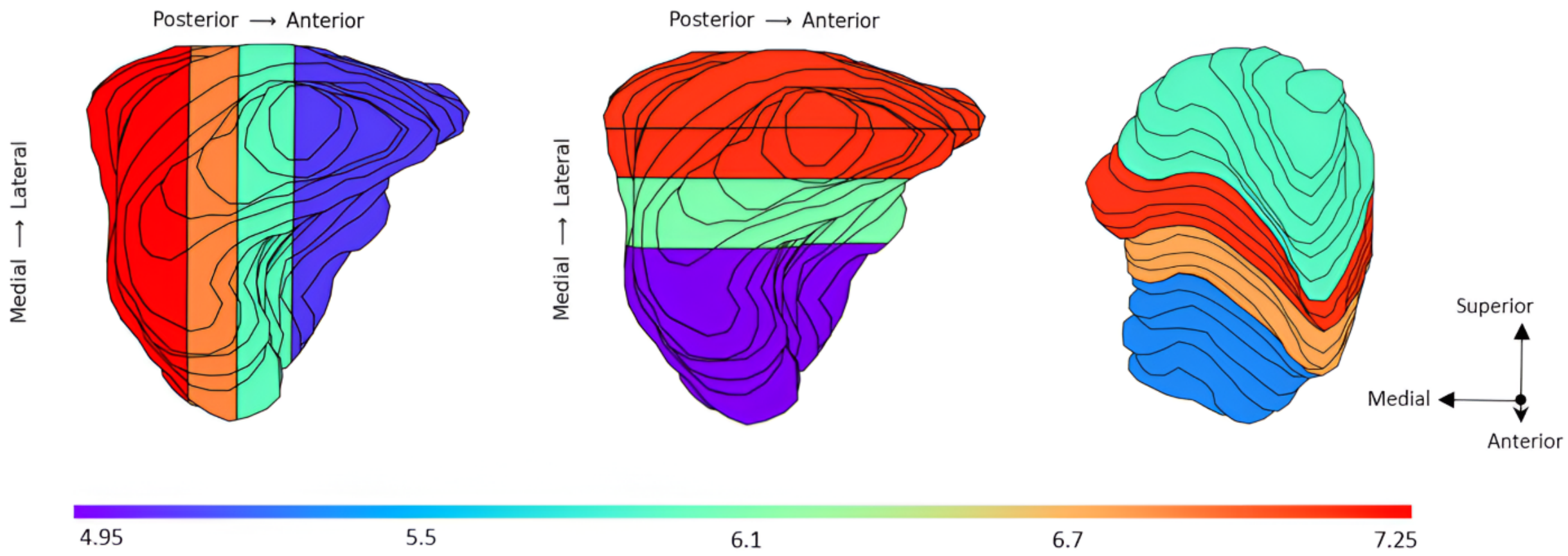

**Fig. 5** $\overline{SUV_{mean}}$ for sub-regions defined by dividing the parotid into four equal-volume regions with parallel planes along each Cartesian direction. The regions of highest uptake for the three cutting planes were the posterior sub-region for the coronal plane cuts (left), the two lateral-most sub-regions for the sagittal plane (middle), and the middle-superior region for the transverse plane (right).

**TABLE 4.** Optimal plane orientation for dividing the parotid gland in half. Orientations were selected to maximize and minimize the difference in $SUV_{mean}$ and $SUV_{max}$ between halves. Uptake statistics on either side of the dividing plane are also provided (i.e. the quantity being minimized and maximized).

| Division Type | | Chopping Plane $(\phi, \theta)$ | $SUV_{above}$ | $SUV_{below}$ |
|---|---|---|---|---|
| $SUV_{mean}$ | Largest Difference | (135°, 90°) | $5.0 \pm 1.5$ | $7.4 \pm 2.3$ |
| | Smallest Difference | (45°, 70°) | $6.1 \pm 1.8$ | $6.2 \pm 1.8$ |
| $SUV_{max}$ | Largest Difference | (165°, 80°) | $13.7 \pm 3.5$ | $15.3 \pm 4.4$ |
| | Smallest Difference | (30°, 40°) | $14.9 \pm 4.2$ | $14.9 \pm 4.4$ |

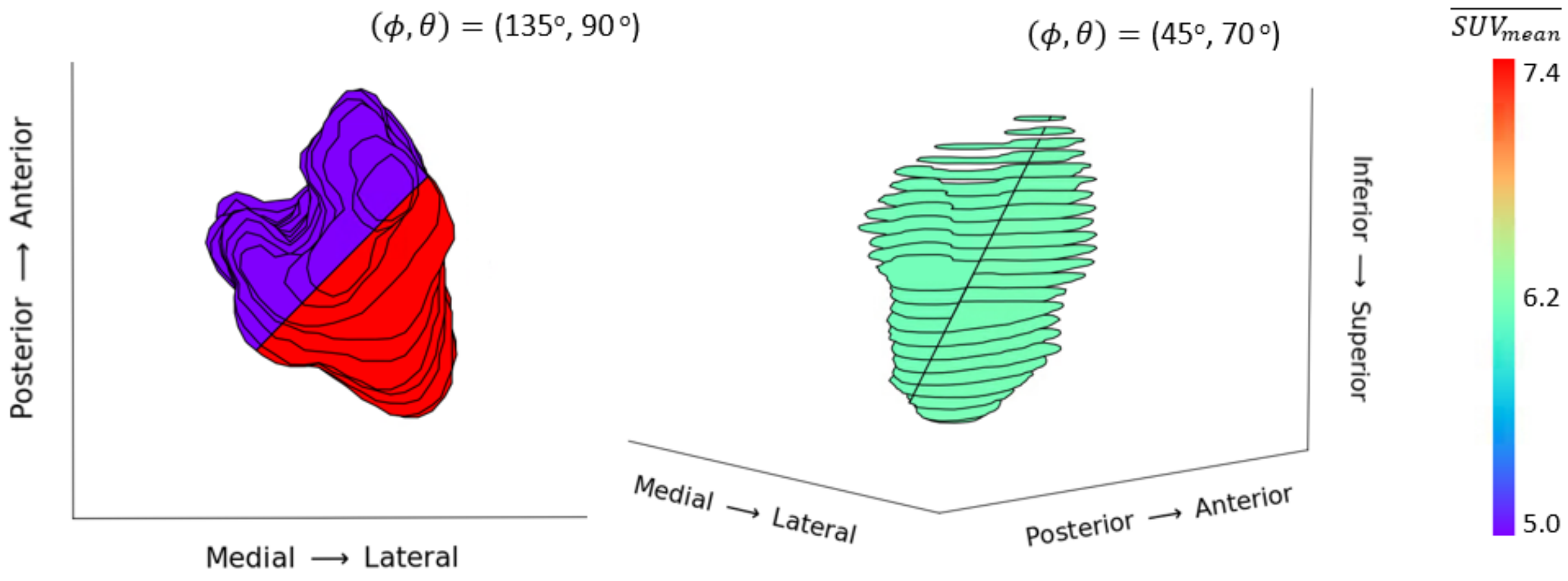

**Fig. 6** The optimal dividing planes for maximizing (left) and minimizing (right) the difference between $\overline{SUV_{mean}}$ in halves of the parotid gland, are illustrated. Planes are defined in spherical coordinates, with the zenith direction in the patient's superior direction, and the azimuthal angle, $\phi$, being 0 in the lateral direction, and increasing towards the posterior direction.

In the 14 analysis sub-regions (4 planar divisions in each Cartesian direction and superficial/deep lobes), the spatial distribution of PSMA PET $\overline{SUV_{mean}}$ was strongly correlated with the GLRLML ($r_s = 0.93, p < 0.001$) and strongly anti-correlated with the GLRLMS ($r_s = 0.94, p < 0.001$). $\overline{SUV_{mean}}$ was not correlated with CT image HUs. These results are summarized in Table 5 and visualized in Fig. 7. Correlations of PSMA PET uptake with GLRLML and GLRLMS remained significant (p < 0.001) following a BH false discovery correction.

Correlations remained strong after re-calculating correlations of uptake and texture features within 18 equal-volume, non-overlapping sub-regions described in the methods ($p < 0.001$). Correlations were approximately unchanged when calculating with either absolute uptake/GLRLM values or values normalized to whole-mean statistics. Whole-gland $SUV_{mean}$ was insignificantly correlated with GLRLML and GLRLMS. These results are summarized in Fig. 8.

**TABLE 5.** Spearman's rank correlation coefficient, $r_s$, and the corresponding p-value for correlations between PSMA PET and CT images inside parotid gland sub-regions. PSMA PET and CT Grey Level Run Length Matrices with long and short run length emphases (GLRLML and GLRLMS) show significant correlations, whereas regular CT Hus do not. Correlations were tested using statistics calculated within the 14 sub-regions considered in the regional uptake analysis, and within 18 non-overlapping, equal-volume sub-regions of parotid glands. Sub-regional statistics were normalized by the whole-mean value before computing correlations.

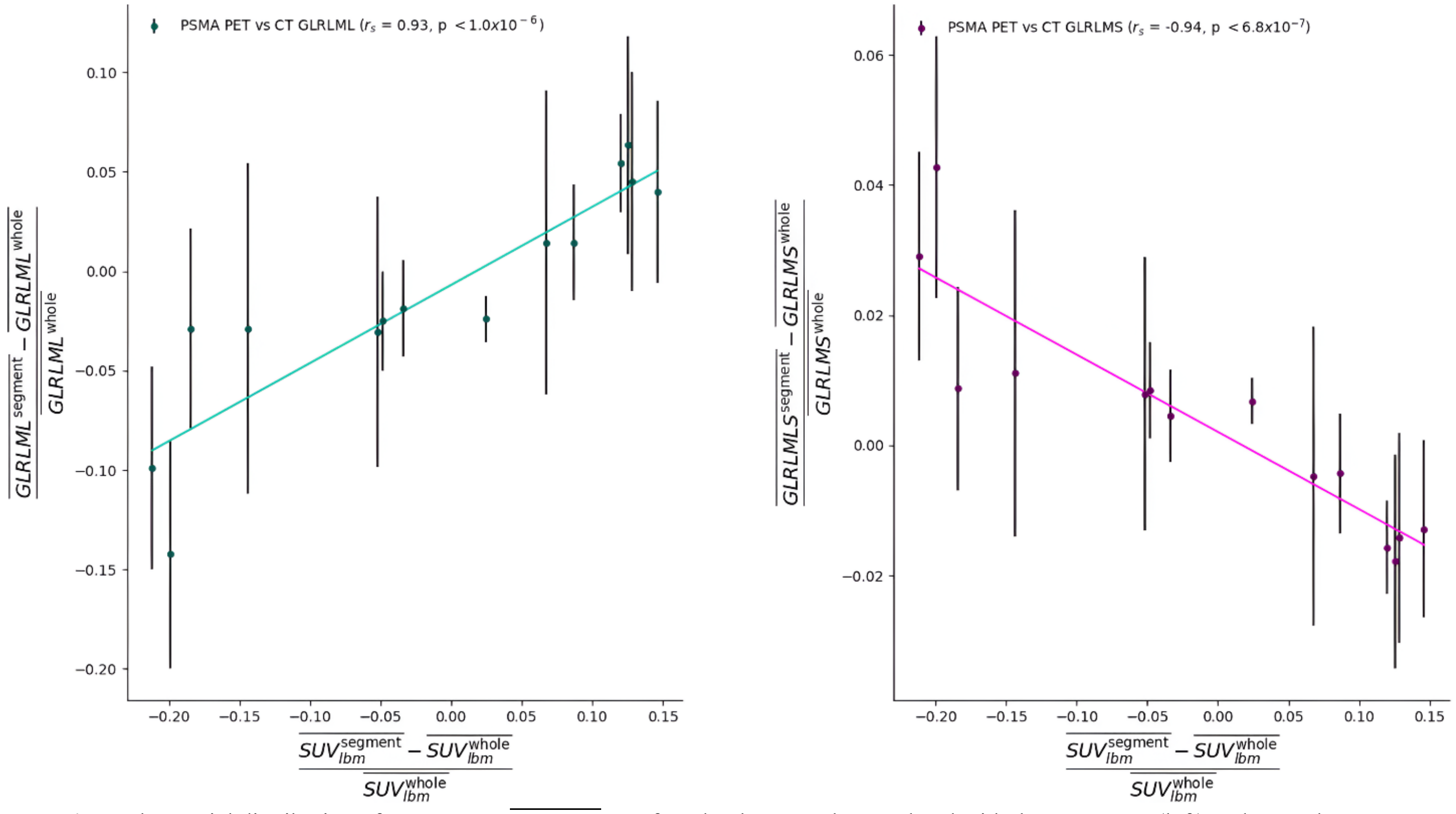

**Fig. 7** The spatial distribution of PSMA PET $\overline{SUV_{mean}}$ was found to be strongly correlated with the GLRLML (left) and strongly anti-correlated with the GLRLMS (right). Correlations were tested using means calculated in 14 sub-regions of parotid glands, represented as normalized differences from whole gland statistics.

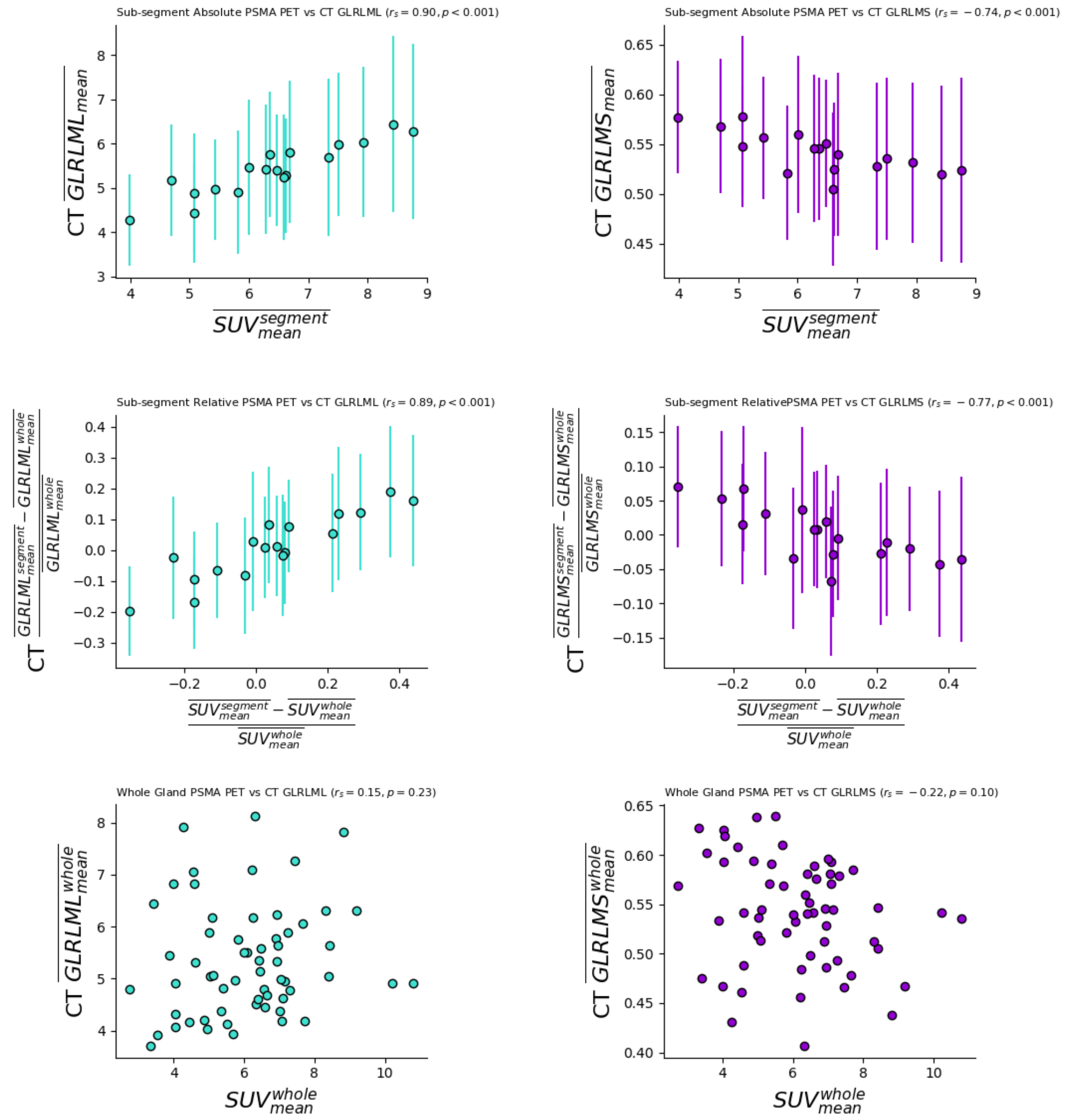

**Fig. 8** Correlations between PSMA PET $\overline{SUV_{mean}}$ and the GLRLML (left) and the GLRLMS (right) were significant when calculated using statistics in 18 non-overlapping, equal volume sub-regions of parotid glands. Correlations were tested using absolute (top) and relative (middle) statistics inside sub-regions. We also tested the correlation between whole-gland $SUV_{mean}$ and CT texture features, over the 60 parotid glands in the dataset. The whole gland statistics were not significantly correlated.

**4 Discussion:**

Spatial heterogeneity of [18F]DCFPyL PSMA PET uptake in the parotid glands has been demonstrated and quantified in well-defined regions. Whole gland statistics agreed with those reported in the literature [7, 26]. The submandibular gland's relatively homogeneous uptake (Fig. 2) and the fact that PET images were attenuation corrected reassured us that regional trends were not simply due to the lateral/posterior regions of glands being closer to the skin surface, or partial volume effects.

In general, $\overline{SUV_{mean}}$ and $\overline{SUV_{max}}$ were found to be highest in sub-regions towards the lateral, posterior, and middle-superior portions of the parotid glands. This was demonstrated by examining sub-regions of the gland defined using SUV thresholds and planar divisions. $\overline{SUV_{mean}}$ was found to increase in an approximately linear fashion from the anterior to the posterior portion of the gland (Fig. 5). $\overline{SUV_{max}}$ did not display as strong of a trend (Table 3). This can be explained by the higher noise associated with a single voxel than the average over a region. In the medial to lateral direction, the $\overline{SUV_{mean}}$ and $\overline{SUV_{max}}$ was significantly higher in the two lateral-most sub-regions than the two medial-most sub-regions. The uptake was found to be more centrally localized in the inferior-superior direction, with the highest uptake in the middle-superior region of the gland.

The superficial lobe was found to have significantly higher PSMA PET uptake than in the deep lobe. The superficial lobe of the parotid gland is situated laterally to the deep lobe and the two lobes are divided by the facial nerve for the purposes of surgical procedures. The two lobes have no previously established anatomical differences [27, 28]. However, superficial parotid lobe-sparing Intensity Modulated Radiation Therapy (IMRT) has been found to significantly reduce the incidence of xerostomia post-RT [29].

While the method of delineating regions of high uptake in the parotid glands via thresholding is favourable for examining patient-specific uptake trends, as it does not require imposing arbitrary divisions, it is challenging for determining the spatial variation of uptake on a population level. Dividing the parotid gland into regions of equal volume using planar divisions lends itself well to cross modality comparisons, as these divisions are well-defined and reproducible without PET images. As PSMA – PET is costly and not routinely acquired for head-and-neck patients, it would be fortuitous if population trends found in PSMA PET uptake were also observable using CT imaging. One such example where CT was used as a proxy for PSMA PET findings is an auto-segmentation model for PSMA PET – defined tubarial glands [30] using only CT images [31].

Determining the best cutting plane for alternatively maximizing and minimizing the difference in $\overline{SUV_{mean}}$ and $\overline{SUV_{max}}$ between halves opens the door to future radiotherapy dose-response studies for validating a correlation or anti-correlation between intra-parotid PSMA PET uptake and functional capacity within the parotid glands. Differences in therapeutic dose between regions with large PSMA PET uptake differences should also lead to differences in xerostomia incidence rates and salivary output measurements. On the contrary, dose differences between regions with minimal differences in PSMA PET uptake could potentially lead to similar patient outcomes. Such a study could also validate the preferential use of $\overline{SUV_{mean}}$ or $\overline{SUV_{max}}$ for characterizing salivary glands.

$\overline{SUV_{mean}}$ and $\overline{SUV_{max}}$ were normalized by lean body mass ($SUV_{lbm}$) as opposed to body weight ($SUV_{bw}$) to reduce the mass-dependence of uptake [32, 33]. Results have been reported in terms of the mean and maximum uptake in regions of the gland, as both metrics tend to be considered in the literature. While the Society of Nuclear medicine and Molecular Imaging (SNMMI) recommends $\overline{SUV_{max}}$ as the preferred tumour uptake metric [34], there is evidence that $\overline{SUV_{mean}}$ has better reproducibility and is less impacted by reconstruction methods [35, 36, 37]. Parotid glands should be expected to have less inter-patient variability than tumour volumes, rendering $\overline{SUV_{mean}}$ to be a suitable metric for quantitative parotid gland imaging with PSMA PET.

The strong correlation of PSMA PET with GLRLML and strong anti-correlation of PSMA PET with GLRLMS implies that regions of high uptake are found to occur in regions of higher homogeneity within CT images (long run lengths). This finding requires further validation. The lack of a correlation between PSMA PET and CT HU images is to be expected, as increased attenuation of X-rays should not be expected from regions of high PSMA PET uptake. Instead, it is the spatial relationships between neighbouring voxels that relate to PSMA PET uptake. Our results suggest that while intra-parotid heterogeneity of PSMA PET and CT texture features appear to be related, the absolute, whole-gland statistics are unrelated. Intra-parotid correlations were approximately unchanged when using absolute statistics or statistics normalized to whole-gland values.

The GLRLM was chosen due to its wide-spread use, simplicity, and interpretability. A short- and long-run emphasis were included since it was unknown which run lengths would have relevance. The number of radiomic texture features tested was minimized to avoid false discovery rates. However, other radiomic texture features, or combinations of features, may have an even better ability to capture PSMA PET uptake trends. A follow-up study focusing solely on correlations between PSMA PET and CT texture features would help to better define these relationships.

The strength of correlations between PSMA PET regional uptake and CT texture features within parotid glands was higher than anticipated. We believe this result is worth emphasizing, as it suggests the possibility of using CT texture features as a proxy for PSMA PET imaging of salivary glands, which is ideal, as PSMA PET images are not expected to become a standard of care for head-and-neck cancer patients. It has been previously demonstrated that PSMA PET uptake patterns may also be linked with T2-weighted magnetic resonance imaging texture features [38].

The relationship between PSMA PET and CT texture features requires further investigation. In particular, it would be of interest to see how uptake patterns of PSMA PET throughout the body correlate with CT texture features, in areas outside the parotid glands. The primary objective of this work was to investigate PSMA PET intra-parotid heterogeneity, and the correlation between PSMA PET and CT texture features should be investigated thoroughly, and throughout the body, in future studies. Furthermore, investigating how intra-parotid PSMA PET uptake relates to gland functionality would facilitate the creation of practical, clinical methods for using CT texture features as a proxy for PSMA PET. We plan to investigate this in a follow-up study.

Computing radiomic features with different voxel sizes is expected to change radiomic feature values. Since our values were averaged over ROIs and only used for computing correlations, we anticipate that the method is robust to the choice of voxel size. It was previously shown that varying slice thickness by 3 mm has a small effect on CT radiomic feature values in liver tumours [39]. However, the voxel size should be considered when comparing results in future works. It may be necessary to standardize voxel sizes for cross-institution radiomic comparisons and analyses. CT was used to delineate parotid glands for this study, and it is possible that some parotid gland tissue was not captured within contours. Variation in delineation strategies between centres will add to variation in reproducibility. Furthermore, reconstruction methods of PET systems and differences between institutions should be considered when reproducing results.

PSMA PET images have increased levels of noise compared to CT, and can be dependent on extrinsic factors, such as food intake [39], which makes the limited cohort (n=30) a weakness of this study. Our results should be externally validated. A retrospective study correlating the dose-response of parotid gland sub-regions with PSMA PET uptake statistics in this study could also assist validation. The results of this study are in line with mounting evidence that parotid glands have a heterogeneous internal anatomy that must be considered for the purposes of radiotherapy planning [40].

In conclusion, PSMA PET uptake in parotid glands is heterogeneous, with regions of high uptake being generally situated towards the posterior and lateral portions of glands. We have identified specific and well-defined regions of high and low uptake in the gland, as well as the optimal planes for dividing parotid glands in half to maximize and minimize differences in PSMA PET uptake. These results lend themselves well to future dose response studies for head-and-neck radiotherapy patients. Lastly, we demonstrated PSMA PET uptake in parotid glands to be strongly correlated with CT texture feature maps, anticipating the potential utility of CT texture feature maps as a proxy for using PSMA PET uptake patterns to tailor patient-specific radiotherapy dose constraints.

**Acknowledgements:**

This work was supported by the Canadian Institutes of Health Research (CIHR) Project Grant PJT-162216.